\documentclass{article}
\usepackage[T1]{fontenc} 
\usepackage[utf8]{inputenc} 
\usepackage{amsmath,cite,url}
\usepackage{amsmath, amssymb, amsfonts, amsthm}
\usepackage{graphicx}
\usepackage{color}
\usepackage{algorithm}
\usepackage{algpseudocode}
\usepackage{mathtools}
\usepackage[margin=1in]{geometry}

\title{Exact, Parallelizable Dynamic Time Warping Alignment with Linear Memory}

  \author{Christopher J. Tralie \\ Ursinus College Dept. of Math/Computer Science \\ \tt ctralie@alumni.princeton.edu \\ Elizabeth Dempsey \\ Ursinus College Dept. of Math/Computer Science \\ \tt eldempsey@ursinus.edu}

\graphicspath{{figs/}}

\newcommand{\warppath}{\ensuremath{\mathcal{W}}}
\DeclarePairedDelimiter\ceil{\lceil}{\rceil}

\newtheorem{theorem}{Theorem}
\newtheorem{lemma}[theorem]{Lemma}

\begin{document}

\maketitle
\begin{abstract}
Audio alignment is a fundamental preprocessing step in many MIR pipelines. For two audio clips with M and N frames, respectively, the most popular approach, dynamic time warping (DTW), has O(MN) requirements in both memory and computation, which is prohibitive for frame-level alignments at reasonable rates. To address this, a variety of memory efficient algorithms exist to approximate the optimal alignment under the DTW cost. To our knowledge, however, no exact algorithms exist that are guaranteed to break the quadratic memory barrier.  In this work, we present a divide and conquer algorithm that computes the exact globally optimal DTW alignment using O(M+N) memory. Its runtime is still O(MN), trading off memory for a 2x increase in computation.  However, the algorithm can be parallelized up to a factor of min(M, N) with the same memory constraints, so it can still run more efficiently than the textbook version with an adequate GPU. We use our algorithm to compute exact alignments on a collection of orchestral music, which we use as ground truth to benchmark the alignment accuracy of several popular approximate alignment schemes at scales that were not previously possible.
\end{abstract}
\section{Introduction}\label{sec:introduction}


The go-to algorithm for computing alignments between two audio clips is Dynamic Time Warping (DTW) \cite{sakoe1970similarity, sakoe1978dynamic}, and DTW and its variants have seen wide application in music processing applications \cite{muller2015fundamentals}.  However, the textbook version of exact DTW has quadratic memory constraints.  While some MIR applications, such as cover song identification, can get away with coarse, beat-synchronous features \cite{ellis2006identifying} to remain in a low memory regime, other applications may require finer scale features and can quickly explode in memory requirements.  For instance, in orchestral music, onsets are weak, so one must often revert to frame-level features for satisfactory alignments.  Singing voices also present unique challenges in this regard \cite{waloschek2018driftin}, and both stringed instruments and singing voices have precise, expressive attacks at sub-beat scales.

In this work, we present a simple divide and conquer variant of DTW to compute a globally optimal alignment between two audio sequences with linear memory.  Our contributions are as theoretical as they are practical; though there are many approximate algorithms that work well in practice (Section~\ref{sec:background}), we are not aware of any other linear algorithms for DTW with this guarantee.  A related advantage is that there are no approximation parameters to tune; there is only one exact cost (with some caveats on numerical precision in Section~\ref{sec:numerical}).

Once we establish the algorithm, we present an experiment on a hand-curated collection of classical music (Section~\ref{sec:experiments}).  Since our memory only scales linearly with a small factor (Section~\ref{sec:memory}), we are able to run it on longer pieces, enabling us to evaluate the precision of approximation algorithms at scales not previously possible.

\section{Background}
\label{sec:background}

\subsection{The Textbook DTW Algorithm}
\label{sec:textbookdtw}

We now briefly review the standard DTW algorithm for context.  Given a (possibly multivariate) time series $X$ with $M$ points and a time series $Y$ with $N$ points, there is a notion of allowable matchings that preserve the time order, known as a time-ordered correspondence, or ``warping path.''  A warping path \warppath, is a correspondence between $X$ and $Y$\footnote{A {\em correspondence} $\mathcal{C}$ between two indexing sets $I$ and $J$ is a subset of the cartesian product $I x J$ so that every element of $I$ is contained in at least one element of $\mathcal{C}$ and every element of $J$ is contained in at least one element of $\mathcal{C}$ } with $K$ ordered tuples of indices of $X$ and $Y$ so that (assuming 0-indexing) $\warppath_1 = (0, 0)$, $\warppath_K = (M-1, N-1)$, and $\warppath_{i} - \warppath_{i-1} \in \left\{ (0, 1), (1, 0), (1, 1) \right\} $.  In other words, matched points between time series must always stay still or advance by at most one along each, and at least one must move forward.

Given a cost measure between the $i^\text{th}$ element $X_i$ in the first time series and the $j^{\text{th}}$ element $Y_j$ in the second time series, $C_{X, Y}(i, j)$, then an ``optimal'' or ``exact'' solution to the {\em Dynamic Time Warping} problem is a warping path $\warppath^*$ that minimizes the sum

\begin{equation}
\sum_{k} C_{X, Y}(\warppath^*_k(1), \warppath^*_k(2))
\end{equation}

We'll refer to an optimal path as $\warppath^*$ and the optimal cost as $D_{X, Y}(M, N)$.  It is possible to compute the  $\warppath^*$ and $D_{X, Y}(M, N)$ using a well-established dynamic programming approach, which is also shared among edit distance algorithms such as Smith Waterman \cite{smith1981identification}.  In particular, if $D_{X, Y}(i, j)$ refers to the optimal cost of aligning the first $i$ points of $X$ to the first $j$ points of $Y$, then the following recurrence holds for $i, j \geq 2$

\begin{equation}
\label{eq:recurrence}
D_{X, Y}(i, j) = \min \left\{  \begin{array}{cc}  D_{X, Y}(i, j-1) & \text{LEFT} \\ D_{X, Y}(i-1, j) & \text{UP} \\ D_{X, Y}(i-1, j-1) & \text{DIAG}  \end{array} \right\} + C_{X, Y}(i, j)
\end{equation}

the boundary conditions are set as 
\begin{equation}
\label{eq:dtwboundary}
D_{X, Y}(0, j) = \sum_{k = 0}^j C_{X, Y}(0, k), D_{X, Y}(i, 0) = \sum_{k = 0}^i C_{X, Y}(k, 0)
\end{equation}

After filling in the first row and column by Equation~\ref{eq:dtwboundary}, it is possible to compute all values of $D_{X, Y}(i, j)$ by applying Equation~\ref{eq:recurrence} from left to right, row by row.  After processing all $MN$ pairs of subsets in this fashion, $D_{X, Y}(M, N)$ contains the optimal cost.  

At this point in the algorithm, we merely have a cost, not an optimal warping path that realizes this cost.  But if we store a second matrix $P(i, j)$ which remembers one of the three ``backpointers'' LEFT, RIGHT, and UP that realized the minimum cost at that cell, then we can ``backtrace'' by following these arrows back from $(M, N)$ to $(1, 1)$ to figure out the elements of an optimal $\warppath$ in between.

\subsection{Variants And DTW Algorithms in MIR}

There are countless works that incorporate and expand on DTW, so we constrain our focus to approaches and conventions that apply to music processing \cite{muller2015fundamentals}, with a particular focus on techniques that accelerate the algorithm.

There is theory to suggest that in general, $O(N^2)$ computation will always be the worst-case for optimal DTW \cite{bringmann2015quadratic}, so many settle for approximate solutions.  The so-called ``Itakura Parallelogram'' \cite{itakura1975minimum} and ``Sakoe-Chiba Band'' \cite{sakoe1978dynamic} were early fixed global alignment restrictions proposed to reduce memory and computation.  More adaptive algorithms have also been used to approximate the DTW alignment on audio streams.  One popular such example is a recursive multiresolution algorithm known as ``FastDTW'' \cite{salvador2004fastdtw}, which has been used to synchronize orchestral music at large scales \cite{muller2006efficient, pratzlich2016memory}.  The algorithm computes the warping path of lower resolutions versions of the time series, and then it recursively constrains alignments at finer scales to lie within some band of the lower resolution path.  It is guaranteed to have worst case $O(M+N)$ run-time and memory consumption.  A similar algorithm, known as ``Memory-Restricted MultiScale DTW'' (MrMsDTW) \cite{pratzlich2016memory} was devised to have constant memory usage, where performance degrades gracefully with a decreasing constant memory, and this technique has proved useful in MIR synchronization applications to pedagogy \cite{tsai2017make}.  We compare to both of these algorithms in Section~\ref{sec:experiments}.  

Beyond this, researchers have cut down on memory with approximate algorithms that use overlapping blocks \cite{macrae2010accurate} and which greedily expand cells to evaluate \cite{dixon2005live}, though, like all of the approximations we've mentioned so far, they have no worst-case approximation guarantees.  The authors of \cite{Ying2016DTW} and \cite{Agarwal2016DTW}, on the other hand, present some of the only algorithms with worst case guarantees for DTW cost in Euclidean spaces.  They achieve linearithmic runtime complexity, with a runtime proportional to the geometric complexity of the time series and inversely proportional to the approximation ratio.  There are also several exact algorithms in the literature that use parallel architectures both for DTW \cite{dijkstra1986minimizing} and for the related problem Smith-Waterman scoring between gene sequences \cite{yu2005smith} to speed up computation.  We draw on these algorithms in our design in Section~\ref{sec:computecost}.  However, they were designed simply to compute costs/scores, not to extract alignments, so we must build on this work to extract alignments (Section~\ref{sec:divideandconquer}).

The closest work in spirit to ours is an algorithm for finding the longest common subsequence between strings \cite{hirschberg1975linear}, which also uses a divide and conquer scheme for sub-problems that overlap on $O(M)$ cells, yielding an algorithm of $O(MN)$ time complexity and $O(M+N)$ space complexity like ours.  However, it does not guarantee a globally optimal solution in the context of DTW \cite{alnaymat2009sparsedtw}.

\section{Our Algorithm}

\subsection{Computing the Cost (DiagDTW)}
\label{sec:computecost}

\begin{figure}
\centering
\includegraphics[width=0.5\textwidth]{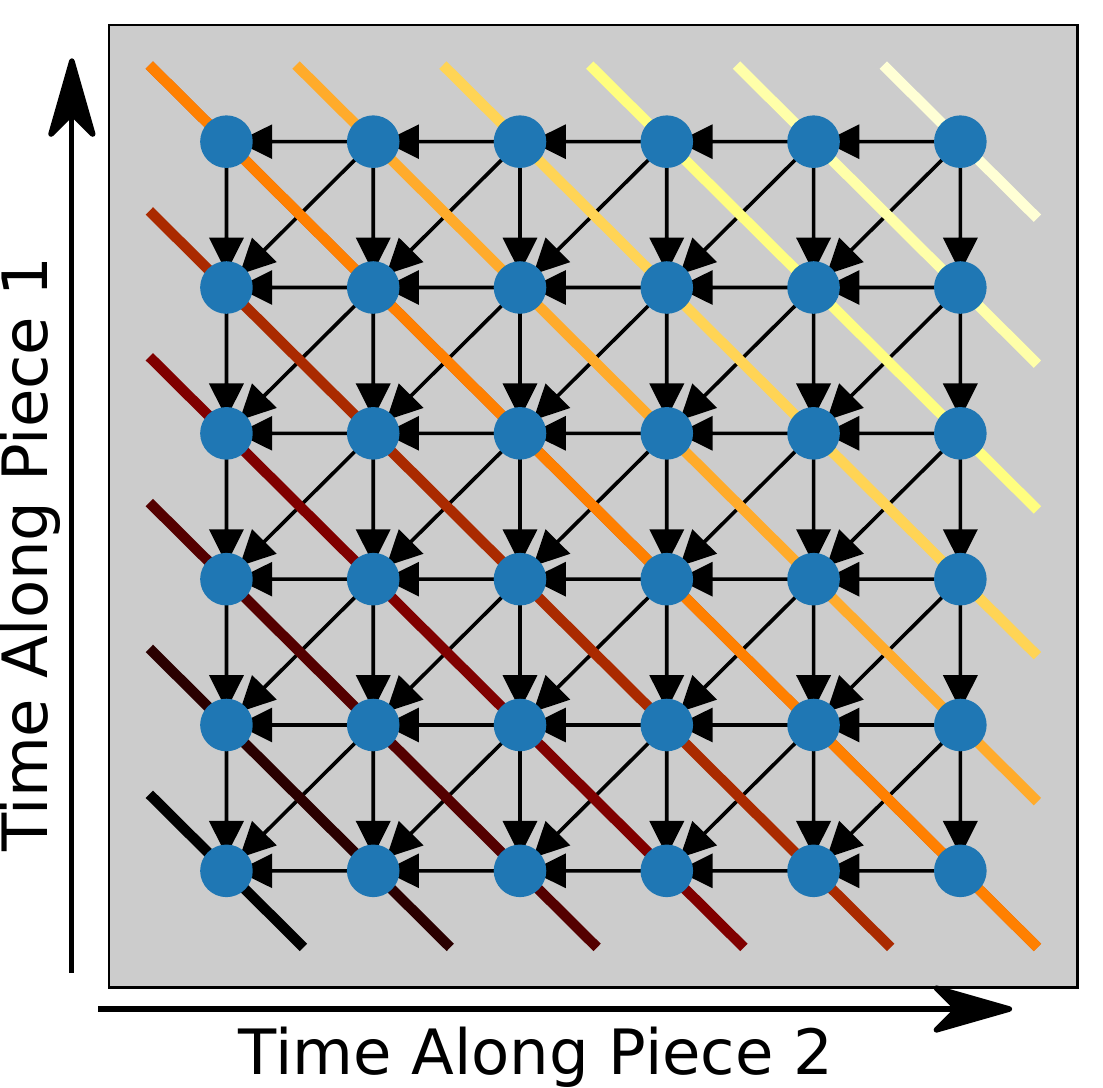}
\caption{A ``linear systolic'' array for computing the DTW cost.  Arrows show dependencies.  All elements along a diagonal can be computed in parallel if the diagonals are processed in order from the lower left (dark) to the upper right (light).}
\label{fig:LinearSystolic}
\end{figure}

The backbone of our algorithm relies on a different order of filling in sub-problems of the alignment, the hardware implementation of which is an instance of a "linear systolic array" in computer architecture \cite{dijkstra1986minimizing, yu2005smith}.  This part is not yet novel, but it is crucial to our approach, so we review it here.  Rather than processing the cells of $D_{X, Y}$ matrix row by row, as in the textbook version, it is also possible to satisfy dependencies needed to complete the recurrences in Equation~\ref{eq:recurrence} while filling in  $D_{X, Y}$ along diagonals.  Figure~\ref{fig:LinearSystolic} shows the idea.  If the diagonals are processed in order from lower left to upper right, then it follows that all elements on a single diagonal $d_k$ can be computed in parallel from two previous adjacent diagonals $d_{k-1}$ and $d_{k-2}$.  Then, to move to the next diagonal, the three diagonal buffers circularly shift; $d_k$ becomes $d_{k-1}$, $d_{k-1}$ becomes $d_{k-2}$, and the previous $d_{k-2}$ can be reused as the new as $d_k$.  Since each diagonal has a max length of $\min(M, N)$, this means that one only needs a memory of $3 \min(M, N)$ to maintain such a system of circularly shifting buffers\footnote{Ignoring the cost of storing features for the moment}; one never needs to store all of $D$ in memory to compute the optimal cost.  Once the algorithm has completed, the optimal cost can be read off as the only element in the last buffer.

\begin{algorithm}
  \caption{Diagonal DTW}\label{alg:dtw}
  \begin{algorithmic}[1]
    \Procedure{DiagDTW}{$X, Y, C_{X, Y},$ kstop}
    \Comment{Time series $X$ and $Y$ with $M$ and $N$ points, a cost $C_{X, Y}$ between them, and the diagonal on which to stop}
    \State $d_0 \gets C_{X, Y}(0, 0)$
    \State $d_1[\gets [C_{X, Y}(1, 0) + d_0[0], C_{X, Y}(0, 1) + d_0[0]]$
    \State $d_2 \gets []$
    \For{$k = 2:$kstop}
        \State Update all elements in $d_2$ based on $d_0$ and $d_1$
        \State $d_0, d_1, d_2 \gets d_1, d_2, d_0$ \Comment{Circularly shift}
    \EndFor \\
    \Return $d_0$, $d_1$, $d_2$
    \EndProcedure
  \end{algorithmic}
  \label{alg:diagdtw}
\end{algorithm}

Algorithm~\ref{alg:diagdtw} shows a sketch of the process.  We refer to this algorithm as {\em DiagDTW}, and we have implemented it on the GPU using CUDA.  For reasons that will become clear in Section~\ref{sec:divideandconquer}, we take as a parameter $2 \leq $ kstop $\leq M+N-1$ on which to stop the computation, and we return the states of all three diagonals at that point.

\subsection{Extracting Alignment}
\label{sec:divideandconquer}

\begin{figure}
\centering
\includegraphics[width=0.7\textwidth]{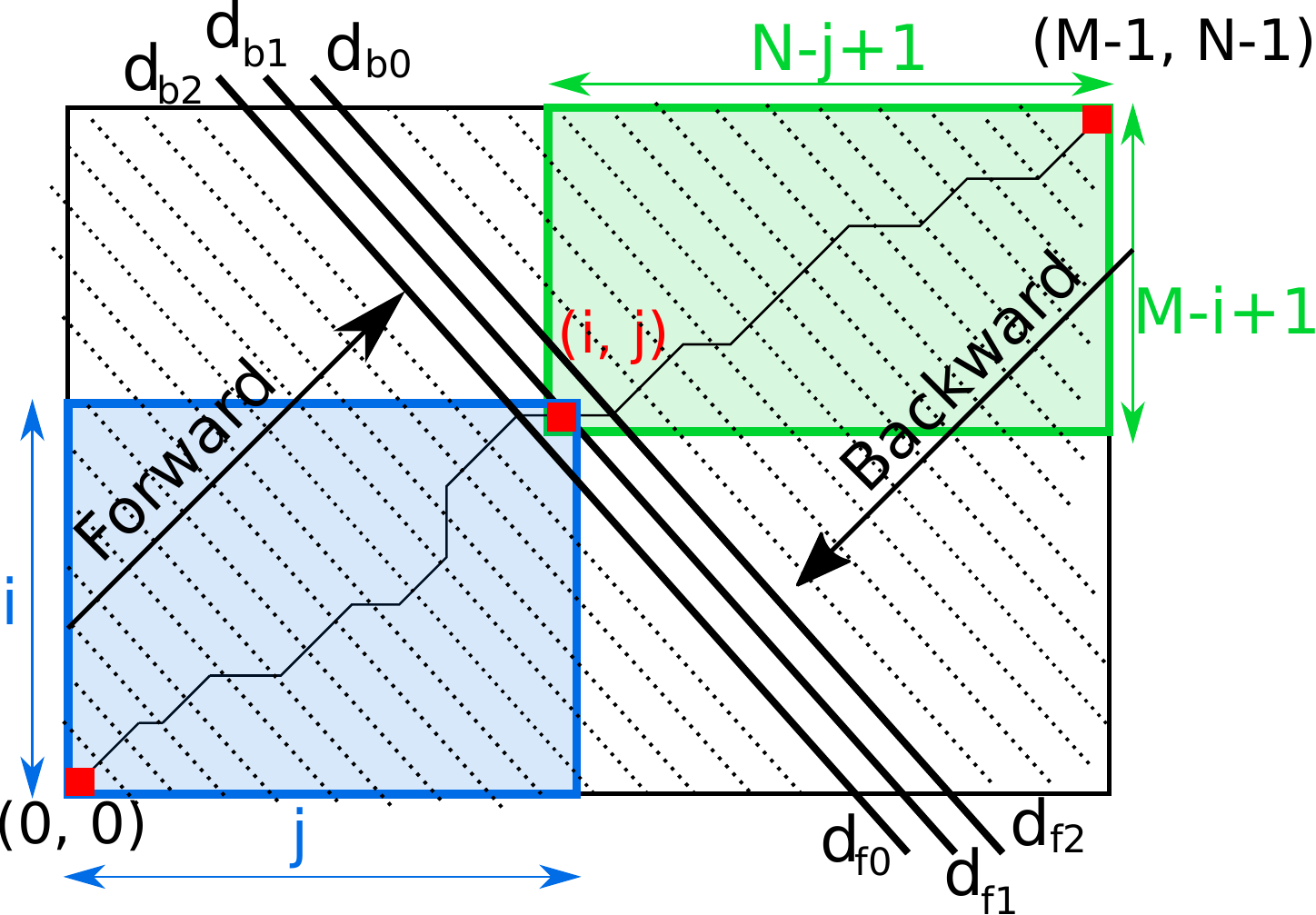}
\caption{Choosing a pivot for the divide and conquer algorithm.  At least one element $(i, j)$ on the optimal warping path $\warppath^*$ resides on one of three central diagonals that overlap from forward and backward computation done halfway, and this element is used to recursively split computation of other warping path points into two halves.}
\label{fig:PathSplit}
\end{figure}

\begin{algorithm}
  \caption{Divide And Conquer Linear Memory Dynamic Time Warping (linmdtw)}\label{alg:dtw}
  \begin{algorithmic}[1]
    \Procedure{linmdtw}{$X, Y, C_{X, Y}, m$}
    \Comment{Time series $X$ and $Y$ with $M$ and $N$ points, cost $C_{X, Y}$ between them, and a minimum dimension $m$}
    
    \If{$M < m$ or $N < m$}\label{line:recursionstop}

        \Return DTW($X, Y, C_{X, Y}$) \Comment{Bruteforce path}
    \EndIf

    \State $K \gets M+N-1$ \Comment{Number of diagonals}
    \State kstop $\gets \ceil*{K/2}$ \Comment{Halfway point}
    \State $d_{f0}$, $d_{f1}$, $d_{f2} \gets$ DiagDTW($X, Y, C_{X, Y}$, kstop) 
    \If{$K$ is even}

        \State kstop $\gets$ kstop $+1$ 
    \EndIf
    \State $X_R \gets \text{reverse}(X)$, $Y_R \gets \text{reverse}(Y)$
    \State $d_{b0}$, $d_{b1}$, $d_{b2} \gets$ DiagDTW($X_R, Y_R, C_{X, Y}$, kstop)
    \Comment{$C_{X, Y}(d_{fk})$ is all costs along the $k^\text{th}$ forward diag}
    \State $d_0 \gets d_{f0} + \text{reverse}(d_{b2}) - C_{X, Y}(d_{f0})$
    \State $d_1 \gets d_{f1} + \text{reverse}(d_{b1}) - C_{X, Y}(d_{f1})$
    \State $d_2 \gets d_{f2} + \text{reverse}(d_{b0}) - C_{X, Y}(d_{f2})$
    \State $(i, j) \gets $ argmin index in $d_0, d_1, d_2$
    \State Now recursively compute other points on $\warppath^*$
    \State LPath $\gets$ linmdtw($X(1,2,..i), Y(1,2,..j), C, m$)
    \State RPath $\gets$ linmdtw($X(i,i+1,..M), Y(j,j+1,..N), C, m$) \\

    \Return LPath + RPath$(2, 3, ...)$ \Comment{Don't double count common point on overlapping sub-paths}
    \EndProcedure
  \end{algorithmic}
  \label{alg:divideandconquer}
\end{algorithm}

The linear systolic array provides a way to compute cost, but if we insist on only remembering the most three recent diagonals of backpointers, then there is no obvious way to recover all of the backpointers to reconstruct an optimal warping path under $O(M+N)$ memory constraints. Instead, we make the following two observations, which we use to build a different algorithm from standard backtracing which works in a memory-restricted setting:

\begin{lemma}
For any warping path $\warppath$ and any adjacent set of 3 diagonals, at least one element of $\warppath$ is incident on one of the three diagonals.
\label{lemma:warppathdiagonal}
\end{lemma}

This follows directly from the definition of a warping path in Section~\ref{sec:textbookdtw}.  We also have the following observation

\begin{lemma}
Let $\warppath^*$ be an optimal warping path and $(i, j) \in \warppath^*$, and let $X_R$ and $Y_R$ be the time series $X$ and $Y$ in reverse order, respectively.  Then the cost of the warping path $\warppath^*$ can be broken into three parts as follows

\begin{equation}
\label{eq:splitcost}
C_{X, Y}(i, j) + C_{X_R, Y_R}(M-i+1, N-j+1) - D_{X, Y}(i, j)
\end{equation}
\label{lemma:splitcost}
\end{lemma}

This is depicted by the overlapping boxes in Figure~\ref{fig:PathSplit}.  In other words, the total cost is the optimal cost of aligning the first half of the path from $(0, 0)$ up to and including $(i, j)$, plus the optimal cost of aligning second half of the path from $(i, j)$ up to and including $(M-1, N-1)$, minus the distance from $X_i$ to $Y_j$ (so we don't double count that distance where they overlap).  This follows from the fact that warping paths must start and end at the beginning and end of each time series (so each sub-path is forced to touch $(i, j)$), and the fact that reversing both time series has no effect on the optimal cost.  This is similar to the observation used in MrMsDTW to break up computation into smaller parts \cite{pratzlich2016memory}.

Now we are ready to present the divide and conquer algorithm to compute an optimal warping path $\warppath^*$.  Lemma~\ref{lemma:warppathdiagonal} and Lemma~\ref{lemma:splitcost} together imply that if we trace the first half of the diagonals in a forward direction and the second half of the diagonals in the reverse direction and add them up pointwise where they meet at the center, subtracting off the distance at those points, then at least one element $(i, j)$ on the three diagonals will contain the optimal cost $C(M, N)$.  Furthermore, since this cost occurs on the optimal path, it will by definition be the {\em minimum cost} over all pointwise sums.  Therefore, to find a point towards the center of $\warppath^*$, we simply do the following
\begin{enumerate}
    \item Run Algorithm~\ref{alg:diagdtw} halfway in the forward direction, starting at the beginning
    \item Run Algorithm~\ref{alg:diagdtw} halfway in the reverse direction, starting at the end
    \item Perform the sums in Equation~\ref{eq:splitcost} where they overlap
    \item Take the indices $(i, j)$ of the value that achieves the minimum over all three diagonals (breaking ties arbitrarily, the result of which we explore in Section~\ref{sec:numerical})
\end{enumerate}.  

We refer to $(i, j)$ as the ``pivot'' at this step, and we are guaranteed that $(i, j)$ resides on $\warppath^*$.  At this point, we divide the problem in half at the pivot and find two more points on the warping path to the left and right, which is the recursive step.  Algorithm~\ref{alg:divideandconquer} summarizes this process.

Because Algorithm~\ref{alg:divideandconquer} calls DiagDTW as a subroutine and DiagDTW uses $3\min(M, N)$ memory, Algorithm~\ref{alg:divideandconquer} also uses at most $3\min(M, N)$ memory.  What is slightly less obvious, but still fairly straightforward to show, is that a serial version of the algorithm takes $O(MN)$ time.  To see this, parameterize the diagonal by a variable $x$, where $x = 0$ at the center of the central diagonal, then the total area of the sub-block to the left of the chosen pivot and to the right of the chosen pivot is bounded from above by the following sum of two products

\begin{align}
A(x) = &\left( \frac{M}{2} + x + 1 \right) \left( \frac{N}{2} - x + 1 \right) \\ +& \left( \frac{N}{2} + x + 1 \right) \left( \frac{M}{2} - x + 1 \right) 
\end{align}

Then, $A'(x) = -4x$, and $A''(x) = -4$, so a global maximum occurs at $x = 0$, for an area of $A(0) = M^2N^2/2 + M + N$.  In other words, ignoring the edge effects $M + N$ due to the overlap, at most half of the total cells are processed across the two halves of each recursive split.  This leads to the recurrence $MN(1 + \frac{1}{2} + \frac{1}{4} + \frac{1}{8} + ...)$, which is bounded from above by $2MN$.  To understand the edge effects, we note the following: since the number of diagonals, $M+N-1$, can be subdivided $\log_2(M+N)$ times, this leads to a bound of $(M + N) \log_2(M+N)$, for a total worst-case cost of 
\begin{equation}
2MN + (M+N) \log_2(M+N)
\end{equation}

However, the $2MN$ term will usually swamp the $(M+N) \log_2(M+N)$, unless one of $M$ or $N$ is very small (e.g $M=1$, in which case it's simply subdividing an interval of length $N$ repeatedly $\log_2(N)$ times).  In practice, we parallelize the DiagDTW step on a GPU, so the algorithm runs faster than this.  We also keep track of the number of cells processed, and we assume $2MN$ to indicate progress of the the algorithm to the user.

Finally, to save the overhead of initiating too many small GPU alignments, we break off the recursion when the sub-blocks get small enough (Line~\ref{line:recursionstop}, Algorithm~\ref{alg:divideandconquer}).  In practice, if either length of of the subdivided time series goes below 500, we use the textbook DTW algorithm to complete the alignment, which uses an insignificant amount of computation and memory at that scale.

\section{Experiments}
\label{sec:experiments}

\begin{table*}[t]
\caption{Memory requirements of the dynamic programming accumulated cost cells for different algorithms on some of the pieces in our dataset.  DTW refers to the naive algorithm, while FastDTW refers to the algorithm in \cite{salvador2004fastdtw} using a band width of $\delta = 30$.  The memory requirements for MrMsDTW for $10^5$ and $10^7$ constant cells is 391KB and 38Mb, respectively}
\begin{tabular}{|l|l|l|l|l|l|}
\hline
Piece & Version 1 & Version 2 & DTW & FastDTW & Ours \\ \hline
Vivaldi Spring & Abbado (188 sec) & Gunzenhauser (209 sec) & 277 MB & 3.86 MB & 194 KB \\ \hline
Candide Overture & Bernstein (268 sec) & Dudamel (279 sec) & 527 MB & 5.5 MB & 270 KB \\ \hline
Beethoven. Symph. No.5 & Thielemann (445 sec) & Bernstein (514 sec) & 1.58 GB & 9.12 MB & 448 KB \\ \hline
Schumann - Symph. No. 3 & Bernstein (2124 sec) & Muti (2199 sec) & 23.2 GB & 36.9 MB & 1.77 MB \\ \hline
Stravinsky The Rite of Spring  & Rattle (2053 sec) & Bernstein (2082 sec) & 29.4 GB & 42.1 MB & 2.02 MB \\ \hline
Tchaikovsky Symph. No. 4 & Bernstein (2645 sec) & Rozhdestve.. (2530 sec) & 46.1 GB & 51.9 MB & 2.48 MB \\ \hline
Shostakovich: Symph. No. 11 & Søndergård (3647 sec) & Nelsons (3765 sec) & 94.6 GB & 74.8 MB & 3.6 MB \\ \hline
Verdi Requiem & Bychkov (4983 sec) & Solti (5042 sec) & 173 GB & 102 MB & 4.9 MB \\ \hline
Wagner - Das Rheingold & Kuhn (8799 sec) & Solti (8759 sec) & 542 GB & 180 MB & 8.6 MB \\ \hline
\end{tabular}
\label{tab:cells}
\end{table*}

\begin{figure}
\centering
\includegraphics[width=\columnwidth]{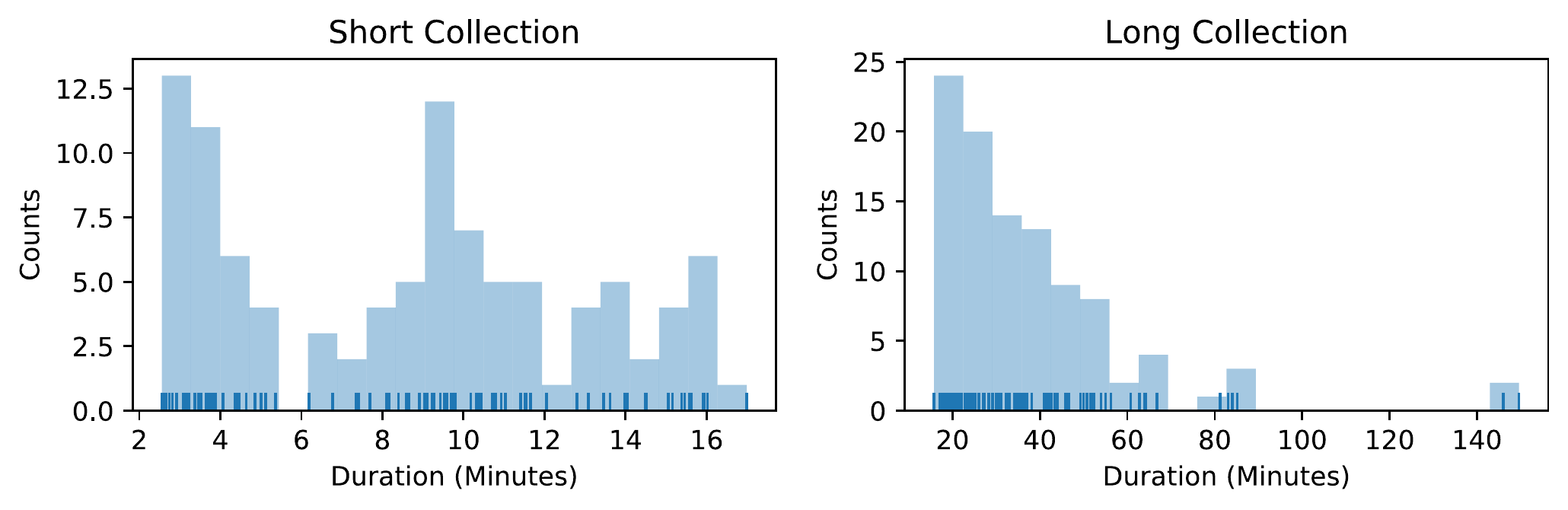}
\caption{The distribution of lengths of our orchestral pieces in the short collection and long collection.}
\label{fig:Counts}
\end{figure}

Now that the theory for our algorithm has been established, we apply it to align real audio data.  We created a dataset with 100 pairs of mostly orchestral pieces, where each pair is performed under different conductors.  All of the performances can be found on Youtube, and we provide code to automatically download them for reproducibility\footnote{If any links go down, the code is robust to that and will simply skip those examples}.  We do not have access to human annotated alignments, but since we can compute exact costs with linear memory, we can use our exact paths to assess the precision of approximate algorithms at very large scales to get an idea of how they perform in that regime.  To that end, we split our dataset into two parts.  The first 50 pairs are ``shorter'' pieces that can be handled (albeit sometimes slowly) by the textbook CPU DTW algorithm.  The second set are pieces that would quickly use up all memory with the textbook algorithm on a personal computer, including many pieces around an hour or longer.  Figure~\ref{fig:Counts} shows the distribution of the lengths of pieces in both sets.

\subsection{Features}

So far, our discussion has assumed that we had access to some distance function $D_{X, Y}$ between time series $X$ and $Y$.  We now finally describe two different features sets that allow us to compute distances for synchronization, which we use in our experiments.  The first set of features are the so-called ``decaying locally adaptive normalized C0'' (DLNC0) features \cite{ewert2009high}, which are popular for fine scale alignments \footnote{Unlike \cite{muller2006efficient}, we do not use a multiscale version of DLNC0, since we are assessing approximations of exact alignments at a single scale}.  The second set of features are referred to as ``mfcc-mod'' features, which consist of a large number of MFCC coefficients, throwing away the lower order ones to control for loudness.  These features were shown to work well at precisely capturing human annotations \cite{gadermaier2019alignmentstudy}.

For both feature sets, we sampled audio at 22050hz, and we used a hop size of 512 between feature frames.  This corresponds to about 43 frames per second of resolution.  For the DLNC0 features, we used librosa's implementation of the CQT with default parameters as a starting point \cite{mcfee2015librosa}.  The DLNC0 features were concatenated to a 0.1 factor of CENS features to improve stability in steady-state regions, as suggested in \cite{ewert2009high}.  For the mfcc-mod coefficients, we used an fft-length of 2048 and computed 120 ``HTK'' coefficients, leaving the first 20 out.  Although \cite{ewert2009high} recommends using cosine distance for the DNLC0 component, we found lower discrepancies using the Euclidean distances as our measure across the board on all of our features.

\subsection{Numerical Precision / Tie Breaking}
\label{sec:numerical}

Since our algorithm is on the GPU, we revert to 32-bit computations, and there is a worry that numerical precision could cause discrepancies, especially since the numbers along warping paths are summed together in a different order in our algorithm, and ties are broken at a different stage.  To rule this out as a source of error when comparing approximation precisions, we compare our GPU answer to the brute force 32-bit CPU answer on the textbook algorithm.  We also compare two different tie breaking rules on 64-bit CPU brute force implementations; one where diagonal takes precedence over left, and one the other way around.  Ultimately, though there are discrepancies, they are negligible compared to errors in approximation, as shown in Figure~\ref{fig:ShorterErrors}.  And the 32-bit versus 64-bit appears to make little difference at these scales.

\subsection{Memory Requirements}
\label{sec:memory}
We compare our alignments to both FastDTW \cite{salvador2004fastdtw} with a band $\delta = 30$ and to MrMsDTW using a constant amount of $10^5$ and $10^7$ cells.  To make Equation~\ref{eq:splitcost} more convenient to compute in our implementation, we store the distances between corresponding points on the three diagonals in addition to the cumulative sums, so we end up using $6 \min(M, N)$ storage instead of $3 \min(M, N)$ storage.  Still, we note that $10^7$ cells is an order of magnitude beyond this requirement, while $10^5$ cells is on the shorter end of what our algorithm needs on the short dataset, so these are two good reference points for MrMsDTW.  To compare memory with FastDTW, we use the equation from \cite{salvador2004fastdtw} which states that the total worst-case space complexity for storing the cells is $N(4\delta + 5)$ values.  Hence, though FastDTW also has linear memory requirements, it has a larger constant factor, particularly for reasonable band sizes ($\delta = 30$ is less than a second of wiggle room).  Table~\ref{tab:cells} shows the memory requirements for storing the cells for different algorithms with variable memory, assuming 32-bit precision (4 bytes per cell).  This neglects the memory for storing the warping path, which is negligible compared to the cost of storing the accumulated cost cells, and it also neglects the memory requirements of storing features, which is a separate issue mostly independent of the algorithms, since these are all run offline.  Still, the memory differences are striking.

\subsection{Results}
\label{sec:results}

\begin{figure}
\centering
\includegraphics[width=0.7\textwidth]{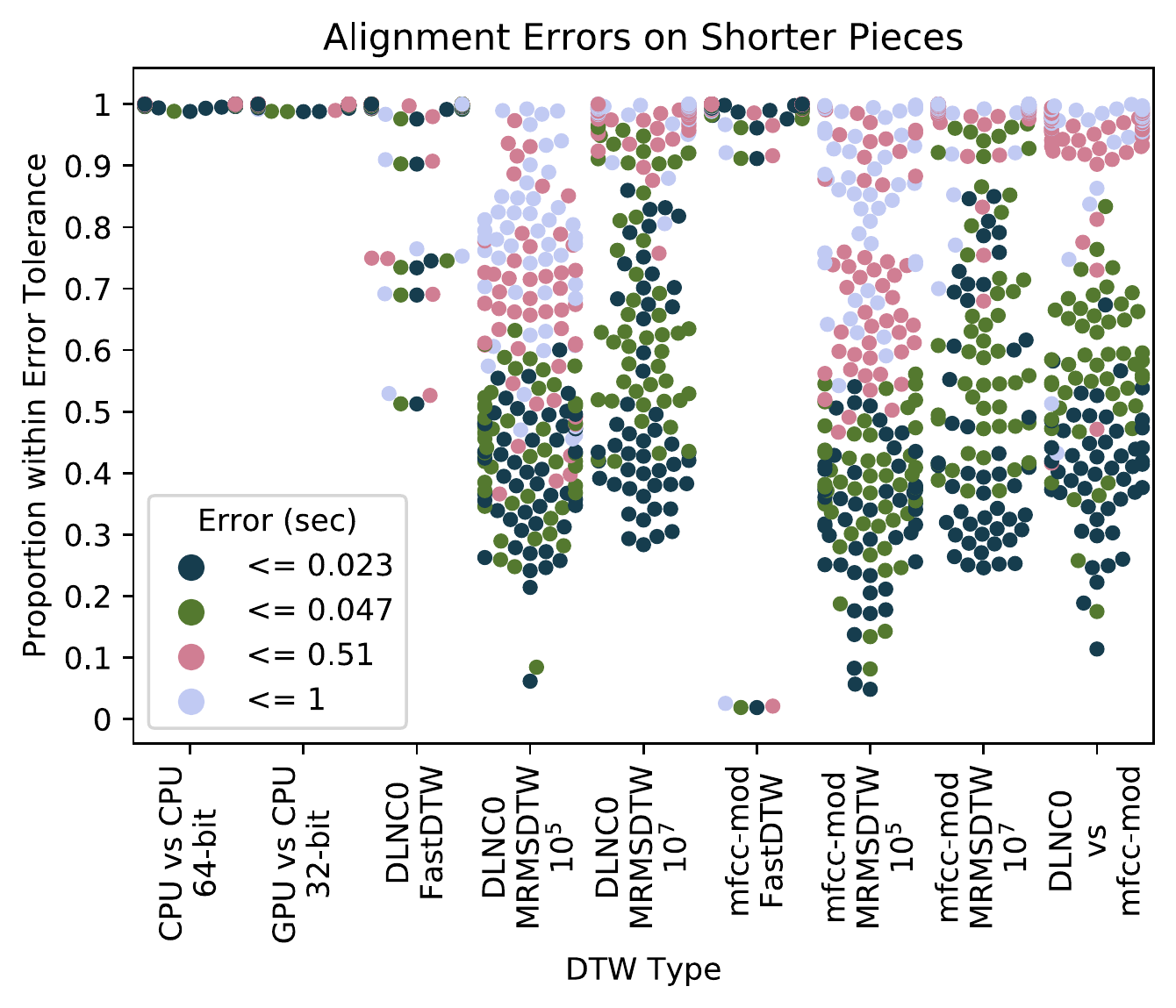}
\caption{Shorter pieces alignment errors.}
\label{fig:ShorterErrors}
\end{figure}

\begin{figure}
\centering
\includegraphics[width=0.7\textwidth]{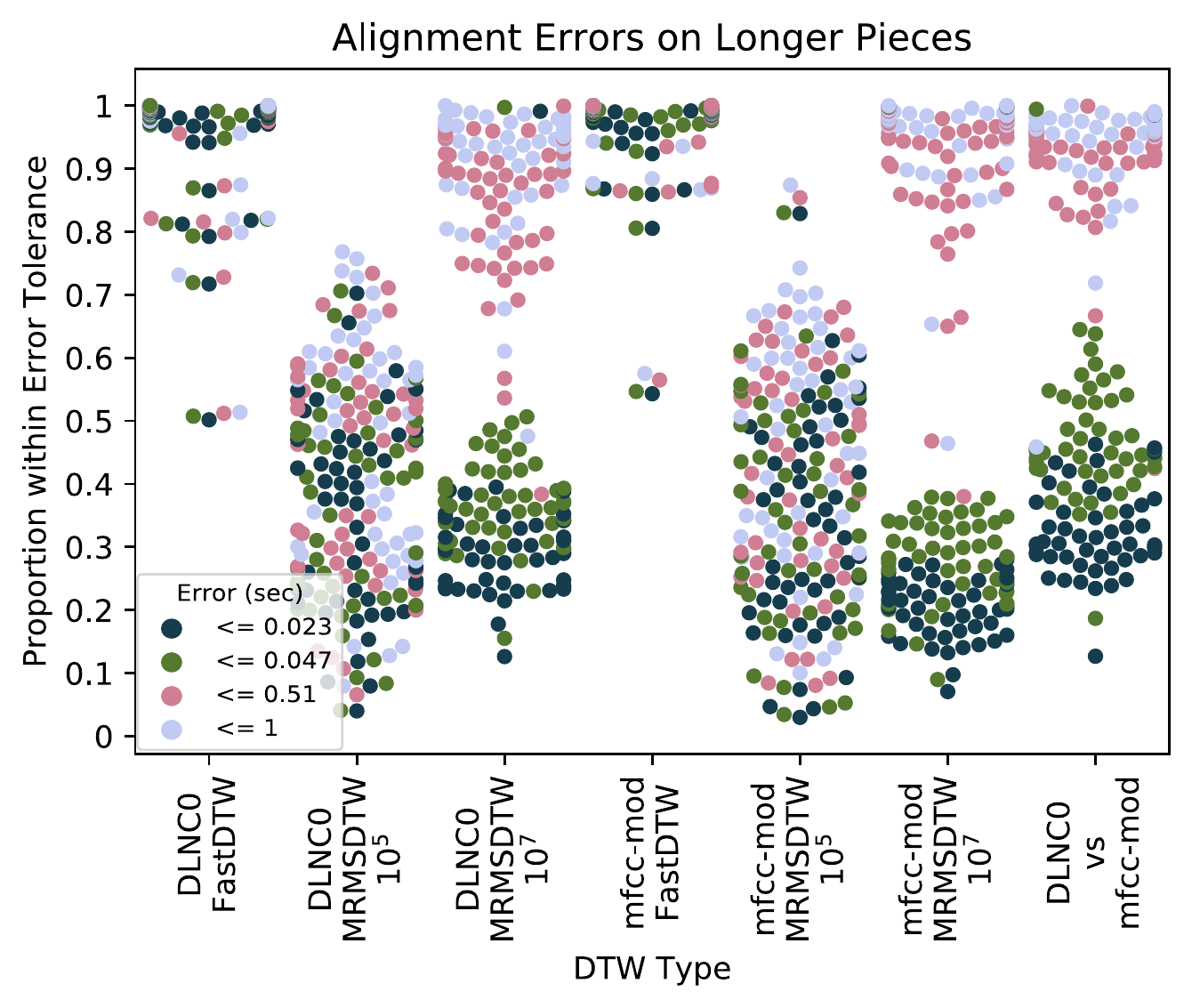}
\caption{Longer pieces alignment errors.}
\label{fig:LongerErrors}
\end{figure}

We now examine the results closely.  We computed the alignment discrepancies between two warping paths $\warppath_1$ and $\warppath_2$ as follows.  For every element $(i, j) \in \warppath_1$, we report the error as $\min |j-k|$ for $(i, k) \in \warppath_2$.  To maintain symmetry, we also add an analogously defined column error to our distributions for every element.  Figure~\ref{fig:ShorterErrors} shows the approximation error distributions for different algorithms on all of the shorter pieces, including tie breaking discrepancies on the exact algorithm (Section~\ref{sec:numerical}), while Figure~\ref{fig:LongerErrors} shows approximation errors on all of the longer pieces.  In each figure for each pairwise comparison, there are four different color dots per piece that indicate the proportion of correspondences $(i, j)$ that fall below the alignment discrepancies (23 ms, 47 ms, 510ms, and 1 second).  Overall, we find that the approximation algorithms often fail to agree at very fine scales, but they usually agree to within a second of audio, which is particularly impressive on the long pieces.  And unsurprisingly, MrMsDTW performs better with more memory.

In addition to approximation errors, we also show the discrepancy between the mfcc-mod and DLNC0 feature sets for reference.  Interestingly, their discrepancy is similar to that of approximation with MrMsDTW, suggesting that feature design is at least as important as a good approximation.  However, under a good feature choice at a fine scale, it is likely that our exact algorithm will give the most desirable alignment.

\begin{figure}[h]
\centering
\includegraphics[width=0.5\textwidth]{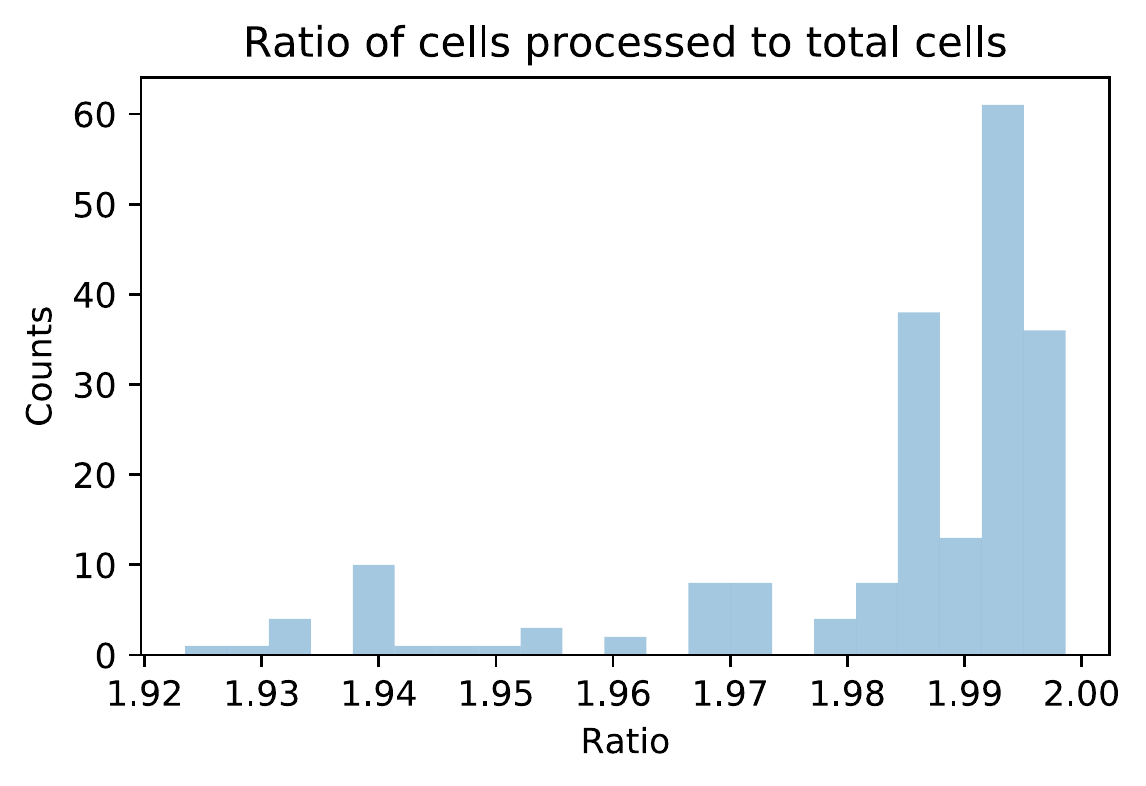}
\caption{In most cases, our algorithm uses close to the factor of 2 bound we established for computation in Section~\ref{sec:divideandconquer}}
\label{fig:Cell}
\end{figure}

Finally, to empirically validate the correctness of our computational complexity bound in Section~\ref{sec:divideandconquer}, we report the ratio of cells processed to total cells in the full accumulated cost matrix in Figure~\ref{fig:Cell} across all pieces, and we find that the ratio is very close to 2 in all cases, as predicted.  Only under very extreme warps away from the center of the matrix would one expect this to be much smaller.

\section{Software}
Since some of the details of linmdtw (Algorithm~\ref{alg:divideandconquer}) are tricky to implement correctly, and in the spirit of reproducibility \cite{mcfee2019open}, we have provided our CPU and GPU (pycuda) implementations of linmdtw, FastDTW, and MrMsDtw in an open source package at \url{https://github.com/ctralie/linmdtw}, which can be installed simply with ``pip install linmdtw''.  We have documentation and Jupyter notebooks on the repo for example usage.  The software will try run CUDA by default, but if it fails, it will fall back to the CPU implementation.  There is also code to replicate the experiments in Section~\ref{sec:experiments} by downloading URLs from Youtube, robustly skipping those no longer available.  Finally, we used the Rubberband Library \cite{cannam2012rubber} and implemented the refinement technique of Ewert (Section 4 of \cite{ewert2008refinement}) to stretch audio to conform to warping paths.

\section{Discussion}

In this paper, we presented a novel exact memory efficient algorithm for DTW.  In addition establishing this new algorithm and proving its correctness, we empirically benchmarked a couple of popular approximation algorithms for DTW alignment in MIR at larger scales than had ever been shown.  We found that these algorithms still have fairly good performance with reference to an exact alignment even on longer pieces.  MrMsDTW is particularly fast computationally, so this suggests that it's good as a first stop in many cases, though there are outliers, and there are always quality gains to be had for an exact algorithm.

Furthermore, though the focus of this paper was on memory constraints, our vanilla GPU implementation also led to speed increases over the textbook CPU version and had similar but slightly slower runtimes than FastDTW.  However, a better GPU implementation would treat global and local memory with more care, along with addressing myriad other issues \cite{xiao2013parallelizing}, so we do not believe this algorithm has yet reached its full computational potential.

There are also other computational problems with very similar dynamic programming design DTW, such as edit distance and Smith Waterman \cite{smith1981identification}, which could benefit from the ability to align large sequences under memory restrictions.  Even approximate DTW algorithms may benefit from tricks in this paper.


\bibliographystyle{plain}
\bibliography{paperarxiv}

\end{document}